\documentclass[twocolumn,floatfix,showpacs,superscriptaddress]{revtex4}

\usepackage{graphicx,amsmath}

\newcommand{\unit}[1]{\,{\rm #1}}
\newcommand{\sub}[1]{_{\mathrm {#1}}}
\newcommand{\subup}[1]{^{\mathrm {#1}}}
\newcommand{\e}[1]{{\rm e}^{#1}}

\newcommand{\vectorsigma}{\boldsymbol {\sigma}}
\newcommand{\vctr}[1]{{\bf {#1}}}
\newcommand{\hc}{{\rm h.\,c.}}
\newcommand{\ann}[1]{#1^{\phantom{\dagger}}}
\newcommand{\cre}[1]{#1^{\dagger}}
\newcommand{\tr}{\rm Tr}
\newcommand{\di}{{\rm d}}

\date{\today}

\begin{document}

\title{Interference of heavy holes in an Aharonov-Bohm ring}

\author{Dimitrije Stepanenko}
\affiliation{Department of Physics, University of Basel,
  Klingelbergstrasse 82, CH-4056 Basel}
\author{Minchul Lee}
\affiliation{Centre de Physique Th\'eorique, UMR6207, Case 907, Luminy, 13288 
Marseille Cedex 9, France}
\author{Guido Burkard}
\affiliation{Department of Physics, University of Konstanz, D-78457
  Konstanz, Germany} 
\author{Daniel Loss}
\affiliation{Department of Physics, University of Basel,
  Klingelbergstrasse 82, CH-4056 Basel}

\begin{abstract}
We study the coherent transport of heavy holes through a
one-dimensional ring in the presence of spin-orbit coupling.
Spin-orbit interaction of holes, cubic in the in-plane components of
momentum, gives rise to an angular momentum dependent spin texture of
the eigenstates and influences transport.  We analyze the dependence
of the resulting differential conductance of the ring on hole
polarization of the leads and the signature of the textures in the
Aharonov-Bohm oscillations when the ring is in a perpendicular
magnetic field.  We find that the polarization-resolved conductance
reveals whether the dominant spin-orbit coupling is of Dresselhaus
or Rashba type, and that the cubic spin-orbit
coupling can be distinguished from the conventional linear coupling by
observing the four-peak structure in the Aharonov-Bohm oscillations.
\end{abstract}


\maketitle

\section{Introduction}

Conductance of mesoscopic rings threaded by the magnetic flux shows
Aharonov-Bohm oscillations \cite{AB59} due to the phase a quantum
state acquires when it winds around the magnetic flux.  An analogous
effect in rings made of semiconductors with spin-orbit coupling
occurs due to the spin precession as an electron orbits the ring,
giving rise to the Aharonov-Casher phase \cite{AC84}.   Both the
Aharonov-Bohm and the Aharonov-Casher effect are manifestations of
quantum coherence in mesoscopic systems, and provide a way to study
the quantum interference in mesoscopic conductors
\cite{LC04,BLK+08,BKS+06,PZ08}.  They lead to universal conductance
fluctuations \cite{UWL+84} and persistent spin and charge currents
\cite{LGB90,LG92,EIA03}.  From a more practical point of view, the conductance
that depends on the magnetic flux in the case of Aharonov-Bohm effect,
or on the strength of spin-orbit coupling in the case of
Aharonov-Casher effect, paves the way for novel applications in
mesoscopic electronic and spintronic devices.  For example, the
Aharonov-Casher phase can be modified by applying a backgate voltage
to the device and changing the Rashba coupling constant \cite{NMT99}.
This enables spintronic devices that require neither any
ferromagnetic materials, nor the control over magnetic field to
operate \cite{NMT99,CRM06,CR07,CR08}.

Recently, a number of experimental \cite{HTS07,GLI+07} and theoretical
\cite{BLK+08} studies have
investigated transport of heavy holes in rings.  These studies
are relevant because of the strong
spin-orbit coupling of heavy holes confined to the ring \cite{GLI+07},
and long coherence length ($\sim 3\unit{\mu
  m}$ in carbon-doped GaAs), making the interference
effects in transport observable.  
%
%
The material parameters of holes
allow for spintronic applications \cite{PGK+04}.  
In coherent spin-orbit coupled
systems, the transport shows an intriguing interplay of Aharonov-Bohm
and Aharonov-Casher effects \cite{EL04}.  Apart from showing strong
spin-orbit coupling and long coherence lengths, the heavy holes
interact through a novel form of the spin-orbit coupling that is cubic
in the in-plane components of momentum.  This form of spin-orbit
coupling influences the interference effects in transport.

In this work, we study the conductance of a ring of heavy holes
tunnel-coupled to two external leads.  This is in contrast to previous
studies which consider rings that are strongly coupled
to the leads \cite{BLK+08}, or are in a diffusive regime and can be
described using semiclassical trajectories \cite{KBJ+07}, or described
in a lattice model \cite{SN04}.  
Studies of the conduction through quantum dots embedded in an
Aharonov-Bohm ring have focused on the effects of interaction on the
transport \cite{HKS01,KG02,SEA05} , while we study the interference of many available paths.
In
these setups, the interference effects can be traced to the
Aharonov-Bohm and Aharonov-Casher phases accumulated by a spin
experiencing a time dependent field while moving along a trajectory
through the ring.  In the adiabatic limit, this approach leads to
geometric phases \cite{LSG93}.  On the other hand, in our tunneling
setup, the quantum effects in transport arise from the interference of
tunneling paths through the eigenstates of the ring.  The
interference is then related to the magnetic field dependence of the
eigenstates of a hole confined to the ring, and not to the phase
accumulated by a spin following quasiclassical trajectory.

The states $|\Psi\sub{hh}\rangle$ of a heavy hole orbiting a ring can
be described in terms of pseudospin textures.  At a position $\phi$
along the ring, the heavy hole state is
\begin{equation}
\label{eq:pseudospin}
\langle \phi | \Psi\sub{hh}\rangle = \psi_+(\phi) |j_z=3/2\rangle + \psi_-(\phi)
|j_z=-3/2\rangle,
\end{equation}
and it determines a unique direction $\vctr{n}$ in the
pseudospin space for which $\langle\phi|\Psi\sub{hh}\rangle$ is an
eigenstate of pseudospin projection to the axis $\vctr{n}$, i.e.,
$|\Psi\sub{hh} \rangle \propto |\sigma_{\vctr{n}}=1\rangle$.
We identify the $|j_z=\pm 3/2\rangle$ heavy hole states with
pseudospin-$1/2$ pointing in $\pm z $ direction,
$|\sigma_{\vctr{e}_z}=\pm 1\rangle$.  The
pseudospin texture associates the direction $\vctr{n}$ with every
point $\phi$ on the ring (Figs.~\ref{fig:textureDresselhaus},
\ref{fig:textureRashba}), so that the states in
Eq.~(\ref{eq:pseudospin}) can be represented in terms of spin texture as
\begin{equation}
\label{eq:deftexture}
\langle \phi | \Psi\sub{hh}\rangle = \e{i\lambda(\phi)}
|\sigma_{\vctr{n}(\phi)}=1\rangle,
\end{equation}
with the texture defined by the position-dependent unit vector
$\vctr{n}(\phi)$, and the position-dependent overall phase $\lambda(\phi)$.  The textures of heavy hole eigenstates
depend on the hole orbital momentum $\kappa$, so that the holes arrive at the
connecting leads with different pseudospins, causing an interference pattern
in the resulting conductance.

In the measurement of conductance as a function of flux through a
semiconductor ring, the Aharonov-Casher effect manifests itself
through an additional structure in the Aharonov-Bohm oscillations due
to spin precession in the arms of the ring \cite{GLI+07}.  In the
approximation of spin-orbit coupling that is linear in momentum, the
conductance oscillations reveal a splitting of Aharonov-Bohm peak in
the Fourier transform of resistivity as a function of the external
magnetic field \cite{YPS02}.  However, the spin-orbit coupling of holes in III-V semiconductors is, in lowest
order, cubic in the hole momentum \cite{BL05}.  In this case, the spin texture
of the orbiting carrier depends on the momentum (see below) and
profoundly influences the transport.  Therefore, for the
carriers with cubic spin-orbit coupling, the Aharonov-Casher phase can
be controlled by changing the momentum of the carriers, without the
need to modify the coupling constant.  This point is especially
important in the structures fabricated in symmetric quantum wells
where the Rashba coupling is absent, and the Dresselhaus spin-orbit
coupling is given by the
crystalline structure.  Even though the coupling constant is fixed, due to the cubic form of spin orbit
coupling, the Aharonov-Casher phase can still be
indirectly controlled through the manipulation of the carrier
momentum.   In addition, the
Dresselhaus and Rashba terms produce different patterns in conductance
as a function of backgate voltage, so that the conductance in
phase-coherent rings reveals the dominant type of spin-orbit coupling.

The remainder of the paper is organized as follows: In Sec. II, we
describe the confinement of heavy holes to a ring and derive the
effective one-dimensional Hamiltonian.  In Sec. III, we solve for the
hole eigenstates and eigenenergies.  In Sec. IV, we introduce the
tunnelling model of hole transport through the ring.  In Sec. V, we
present the resulting differential conductance of the ring.  We conclude in Sec. VI. 
 
\section{Heavy holes in a one-dimensional ring}

Heavy holes confined to the two-dimensional hole gas (2DHG) are described with $H = H_0
+ H\sub{SO} + H\sub{Z}$, where $H_0={\vctr{p}^2}/2m\sub{hh}$ is
the standard kinetic term,
$H\sub{Z}=(1/2)\mu\sub{B}\vctr{B}\cdot {\bf g} \cdot \vectorsigma$ is the
Zeeman coupling to the magnetic field $\vctr{B}$, $\mu\sub{B}$ being
the Bohr magneton, ${\bf g}$ the gyromagnetic tensor of the
confined holes, and $\vectorsigma$ the vector of the pseudospin Pauli
matrices.  The spin-orbit interaction of heavy holes is, in
lowest order, cubic in the in-plane components of the momentum \cite{BL05},   
\begin{equation}
\label{eq:hso}
H\sub{SO} = \left( i\alpha p_-^3 + \beta p_-p_+p_- \right)\sigma_+ + \hc,
\end{equation}
where $\alpha$ and $\beta$ are respectively interaction strengths of
Rashba and Dresselhaus spin-orbit coupling, and $O_{\pm}=O_x \pm i O_y$,
($O=p,\sigma$).  The pseudo-spin represents the two heavy hole states
$|\sigma_z=\pm1\rangle=|j=3/2, j_z=\pm 3/2\rangle$.  This is in sharp
contrast to the electrons in a two-dimensional electron gas (2DEG),
where the spin-orbit is in the lowest order linear in momentum.  Effects of spin orbit coupling in general
depend on the confinement, both to the 2DHG and to the ring.  We will
treat the spin-orbit coupling strengths $\alpha$ and $\beta$ as
free parameters and absorb the influence of the electrostatic potential that
confines the holes to two dimensions into their values.  In particular,
if the confinement to two dimensions is caused by a symmetric
potential, the Rashba coupling vanishes, $\alpha =0$.  We neglect
the orbital effects of the magnetic fields so that $\vctr{p}$ is the
kinetic momentum of the hole.  

In order to illustrate the spin structure of ring eigenstates, we will
first solve for the eigenvalues and wave functions of the heavy holes
confined to the ring in the absence of magnetic field.  Later, we take
the magnetic field into account and find that it causes modification
of the quantization condition and the Zeeman coupling.

The two-dimensional hole gas is confined to the ring by a radial potential
$V(r)$ that has a deep minimum in the interval $a-w/2<r<a+w/2$, where
$a$ is the radius of the ring, and $w$ is its width.  States of the
hole orbiting the ring in the limit of strong confinement are products
of the ground state radial wave function in the potential
$V(r)$, and a function of the angular coordinate $\Psi(\phi)$.  For
strong radial confinement, the motion of the hole in the ring is
described by an effective Hamiltonian that depends only on the angular
coordinate along the ring and all the properties of the radial wave
function enter the problem only through the parameters of the
effective one-dimensional Hamiltonian.  The description in terms of
the effective one-dimensional Hamiltonian is valid when both the
energy spacing of the 2DHG confinement and the energy spacing of the
radial confinement are much larger than the energies associated
with the motion along the ring.

We find an effective Hamiltonian for the ring by introducing the
confinement potential $V(r)$ in the radial direction in $H\sub{SO}$
and reducing it to the subspace of the lowest radial mode, in analogy
with \cite{MMK02}.  Typically these results were obtained by
introducing a model potential and explicitly calculating the angular
Hamiltonian for the lowest radial mode.  The resulting one-dimensional
effective Hamiltonian for the harmonic radial confinement was found for
the case of linear- \cite{MMK02} and cubic Rashba \cite{KBJ+07} spin-orbit coupling.  We note that
generically the solution to the radial problem in an arbitrary
potential can lead to divergences in the effective Hamiltonian.  This
can be avoided by working directly with the radial wave function in the
form of a harmonic oscillator ground state.  In this work, we employ a
different approach, and calculate the effective Hamiltonian for a
general radial wave function.  The resulting effective one-dimensional Hamiltonian is
\begin{equation}
\label{eq:h1d}
\begin{split}
H = &-\frac{1}{2m\sub{hh}a^2}\partial^2_{\phi} + 
\\
&
\left[ i\alpha \e{3i\phi}\left(F_0 + F_1 \partial _\phi + F_2 \partial ^2 _\phi
  + F_3 \partial ^3_\phi\right) + \right.
\\
& \left. ~ \beta \e{i\phi}\left( G_0 +G_1 \partial_\phi +G_2 \partial ^2 _\phi +G_3
\partial ^3 _\phi \right)\right] \sigma_-  +  
\\
&
\left[ -i\alpha \e{-3i\phi}\left(F_0 - F_1 \partial _\phi + F_2 \partial ^2 _\phi
  - F_3 \partial ^3_\phi\right) + \right.
\\
& \left. ~ \beta \e{-i\phi}\left( G_0 - G_1 \partial_\phi + G_2 \partial ^2
_\phi - G_3 \partial ^3 _\phi \right)\right] \sigma_+.
\end{split}
\end{equation}
where $G_0=i(R_0+R_1-R_2)$, $G_1=-(R_1+R_2)$, $G_2=i(R_2-2R_3)$, and
$G_3=-R_3$, $F_0 = i(R_0 - 3 R_1 + 3 R_2)$, $F_1 = -3R_1 + 9R_2 - 8
R_3$, $F_2 = i(-3R_2 + 6R_3)$, and $F_3=R_3$.  The parameters that
depend on the radial confinement are $R_j = \langle
r^{-j}\partial^{3-j}_r \rangle\sub{radial}$, where the expectation
value is taken in the ground state radial wave function.    The
parameters $R_j$, $j=0,\ldots,3$, satisfy consistency conditions that
reduce the number of free parameters to two.  We keep the explicit
dependence of the independent expectation values in the radial state.
The constraints are $R_2=R_3/2$ and $R_0=-3R_1/2$.  The constraints
can be proven using integration by parts in the radial part of the
Schr\"odinger equation, under the assumption that the radial part of the
wave function vanishes at the origin together with its derivatives up
to order 3.  We have checked that this conclusion holds in the limit
of a series of potentials that converge to the hard wall.  Also, note
that the relation between $R_0$ and $R_1$ is satisfied for the radial
wave functions of the harmonic confinement for which $R_3$ diverges
\cite{KBJ+07}.  We can take the values $R_0$ and $R_3$ as the free
parameters of the ring confinement.  For a ring of radius $a$ and
width $w$, $R_3\propto a^{-3}$, $R_0\propto a^{-1}w^{-2}$.  

Before embarking on the solution of the one-dimensional problem, let us briefly
discuss the resulting Hamiltonian.  Depending on the radius and the
width of the ring, different terms in the spin-orbit interaction
become more or less important.  Also, we see an enhancement of the
spin-orbit effects in narrow and small rings.   We see that the
strength of the spin-orbit coupling terms depends on the width of the
one-dimensional ring $w$ through the parameter $1/(aw^2)$.  This means that the
spin-orbit coupling terms can be enhanced in a very narrow ring.  In
this limit, however, the spin-orbit coupling is effectively linear.
Therefore, the effects of the cubic spin-orbit coupling presented here
will be pronounced in the rings of intermediate widths, and the
strength of radial confinement that is strong enough for the
approximation of the single radial mode to hold.

\section{Spectrum and eigenstates of the orbiting holes} 

The effective Hamiltonian (\ref{eq:h1d}) describes a ring of heavy
holes in the presence of both Dresselhaus and Rashba spin-orbit
interaction, when $\alpha \neq 0$ and $\beta \neq 0$.  Our goal is to
understand the role of cubic spin-orbit coupling in transport, and
contrast its effects to the standard linear spin-orbit coupling,
experienced by the electrons in a similar configuration.  We will
therefore focus on the two limits that allow for a simple solution,
namely, Dresselhaus-only interaction ($\alpha = 0$), and Rashba-only
interaction ($\beta=0$) that was previously studied in \cite{KBJ+07}.
While restricting the domain of validity of our results, these
approximations emphasize the physical picture of the eigenstates in
terms of holes orbiting the ring, and the associated texture of the
hole pseudospin.  Apart from allowing a simple solution and providing
a simple picture of the eigenstates, these two limits are also, in
principle, realizable in practice.  In the semiconductor
heterostructures that confine holes to the 2DHG, the strength of the
Rashba term is governed by the asymmetry of the confining potential in
the direction perpendicular to the 2DHG plane.  For a highly
asymmetric potential the Rashba term is dominant, but it vanishes when
the holes are confined by a symmetric potential well.

\subsection{Dresselhaus ($\alpha = 0$) case}
 
\begin{figure}[t]
\begin{center}
\includegraphics[width=8cm]{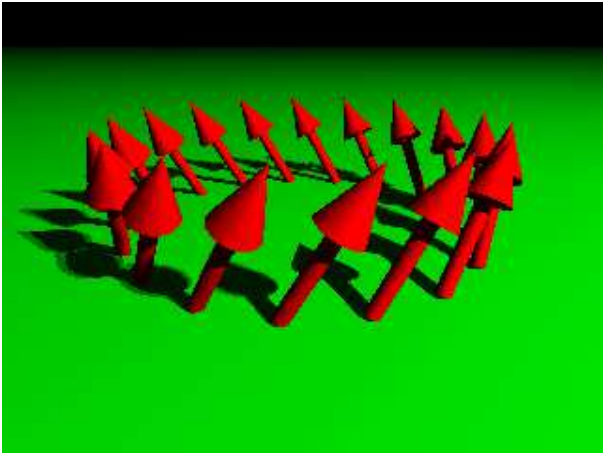}
\caption{\label{fig:textureDresselhaus} (Color online) Hole pseudospin
texture of the Dresselhaus-only eigenstate.}
\end{center}
\end{figure}

Eigenstates of the effective Hamiltonian (\ref{eq:h1d}) are specified
by two quantum numbers, $\kappa = (2n +1)/2$, where $n$ is an integer,
and the texture quantum number $\tau=\Uparrow, \Downarrow$, that takes
on two discrete values.  The Dresselhaus interaction eigenstates
$\Psi\subup{d}$ are 
\begin{align}
\label{eq:kappaUpDresselhaus}
\Psi_{\kappa\Uparrow}\subup{d}&=\e{i\kappa\phi}
\left(
\begin{array}{c}
\cos\frac{\theta\subup{d}(\kappa)}{2}\e{-\frac{i}{2}(\phi + \frac{\pi}{2})}
\\
\sin\frac{\theta\subup{d}(\kappa)}{2}\e{\frac{i}{2}(\phi + \frac{\pi}{2})}
\end{array}
\right)
\\
\label{eq:kappaDownDresselhaus}
\Psi_{\kappa\Downarrow}\subup{d}&=
\e{i\kappa\phi}
\left(
\begin{array}{c}
-\sin\frac{\theta\subup{d}(\kappa)}{2}\e{-\frac{i}{2}(\phi + \frac{\pi}{2})}
\\
\cos\frac{\theta\subup{d}(\kappa)}{2}\e{\frac{i}{2}(\phi + \frac{\pi}{2})}
\end{array}
\right),
\end{align}
where the texture angle $\theta\subup{d}(\kappa)$ is
\begin{equation}
\label{eq:textureAngleDresselhaus}
\theta\subup{d}(\kappa) = \tan^{-1}
\left[\frac{2m\sub{hh}\beta}{R_3^{2/3}} \left( \frac{2}{3}R_0 + 
\left(\kappa^2 - \frac{5}{4}\right)R_3\right)\right].
\end{equation}
The states represent a hole that orbits the ring with angular momentum
$\kappa$ and well defined spin texture.  At the point on the ring with
the angle $\phi$, the spin state
$\langle\phi=\phi_0|\Psi\subup{d}_{\kappa\Uparrow}\rangle$ corresponds
to the spin that is tilted by the angle $\theta\subup{d}(\kappa)$ from
the normal to the plane of the ring, and the azimuthal angle is $\Phi = \phi_0 +
\pi/2$, so that the projection of the spin to the plane of the ring is
always tangential to the ring (see Fig.~\ref{fig:textureDresselhaus}).  The spin state associated with the
other texture, $\langle\phi=\phi_0|\Psi_{\kappa\Downarrow}\rangle$,
corresponds to the spin with the tilt angle $\pi - \theta\subup{d}$,
and the same azimuthal angle.  The crucial difference with respect to
the eigenstates of the ring with linear spin-orbit coupling is that
the texture of the state depends on the momentum quantum number
$\kappa$ even in the absence of magnetic fields.  Therefore, the
states of different momentum show different spin textures.

Energies depend on both momentum and the texture,
\begin{equation}
\label{eq:energiesDresselhaus}
E_{\kappa,\Uparrow(\Downarrow)}\subup{d} =
\frac{1}{2m\sub{hh}R_3^{2/3}}\left( \kappa^2 + \frac{1}{4} \pm
\frac{\kappa}{\cos\theta\subup{d}(\kappa)}\right).
\end{equation} 
The pairs of eigenstates
$(\Psi_{\kappa,\Uparrow},\Psi_{-\kappa,\Downarrow})$ form Kramers
doublets, $E\subup{d}_{\kappa \Uparrow}= E\subup{d}_{-\kappa \Downarrow}$.

\subsection{Rashba ($\beta = 0$) case}

\begin{figure}[t]
\begin{center}
\includegraphics[width=8cm]{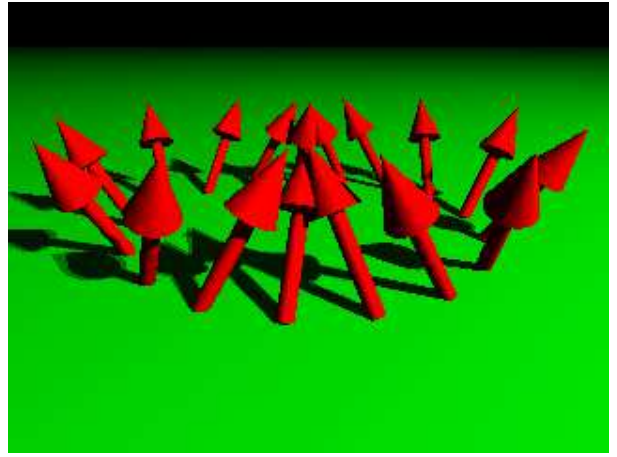}
\caption{\label{fig:textureRashba} (Color online) Hole pseudospin
texture of the Rashba-only eigenstate.}
\end{center}
\end{figure}

When the spin-orbit coupling is of the Rashba type, the momentum
$\kappa = (2n+1)/2$ for integer $n$ is still a good quantum number, and
there are still two textures, $\tau=\Uparrow$ and $\tau=\Downarrow$
for every value of $\kappa$.  The eigenstates $\Psi\subup{r}$ are
\begin{align}
\label{eq:kappaUpRashba}
\Psi_{\kappa\Uparrow}\subup{r}&=\e{i\kappa\phi}
\left(
\begin{array}{c}
\cos\frac{\theta\subup{r}(\kappa)}{2}\e{-\frac{3i}{2}\phi}
\\
\sin\frac{\theta\subup{r}(\kappa)}{2}\e{\frac{3i}{2}\phi}
\end{array}
\right)
\\
\label{eq:kappaDownRashba}
\Psi_{\kappa\Downarrow}\subup{r}&=
\e{i\kappa\phi}
\left(
\begin{array}{c}
-\sin\frac{\theta\subup{r}(\kappa)}{2}\e{-\frac{3i}{2}\phi}
\\
\cos\frac{\theta\subup{r}(\kappa)}{2}\e{\frac{3i}{2}\phi}
\end{array}
\right),
\end{align}
with the Rashba texture angle $\theta\subup{r}(\kappa)$
\begin{equation}
\label{eq:textureAngleRashba}
\theta\subup{r}(\kappa) = \tan^{-1} \left[\frac{2m\sub{hh}
\alpha}{R_3^{2/3}} \left( \frac{2}{3}R_0 +
\left(\frac{13}{12}-\frac{1}{3}\kappa^2\right)R_3\right)\right].
\end{equation}
As for the Dresselhaus case, the eigenstates represent a hole with
well defined pseudospin texture that orbits the ring.  The texture is
however quite different.  The pseudospin $\langle \phi = \phi_0|
\Psi\subup{r}_{\kappa \Uparrow}\rangle$ is tilted away from the normal to
the ring plane by the angle $\theta\subup{r}(\kappa)$ that, in
contrast to the Dresselhaus case, can vary in the full range
$\theta\subup{r}\in [0,\pi]$, while the Dresselhaus spin orbit
coupling allows only for $\theta\subup{d} \in [0,\pi/2]$, except for
$\kappa=1/2$ and unrealistically large $R_3$.  The pseudospin projection to the
plane of the ring, that was always tangential in the Dresselhaus case, now
makes three full rotations on each orbit (see Fig.~\ref{fig:textureRashba}).  The pseudospin of the
opposite texture $\langle \phi = \phi_0| \Psi\subup{r}_{\kappa
  \Downarrow}\rangle$ has the tilt angle $\theta = \pi -
\theta\subup{r}(\kappa)$, and the same projection to the ring plane.

Energies in the Rashba case again depend on the momentum and texture
\begin{equation}
\label{eq:energiesRashba}
E_{\kappa,\Uparrow(\Downarrow)}\subup{r} =
\frac{1}{2m\sub{hh}R_3^{2/3}}\left( \kappa^2 + \frac{9}{4} \pm
\frac{\kappa}{\cos\theta\subup{r}(\kappa)}\right).
\end{equation} 
The time reversal symmetry imposes Kramers degeneracy, and the
states in the Kramers doublet $(\Psi\subup{r}_{\kappa
  \Uparrow},\Psi\subup{r}_{-\kappa \Downarrow})$ have the same energy
$E\subup{r}_{\kappa \Uparrow} = E\subup{r}_{-\kappa\Downarrow}$. 

\subsection{Magnetic field}

Our preceding calculation of the eigenstates and eigenenergies did not take
into account the interaction of holes with the magnetic field $\vctr{B}$.  In
this subsection, we will find the spectrum and the eigenstates of a heavy
hole in the presence of a magnetic field normal to the ring.  This
calculation includes the change of the quantization condition for the
orbital momentum $\kappa$ and the Zeeman term $H\sub{Z}$, but neglects
the modification of the lowest energy radial wave function due to the
magnetic confinement.  This approximation neglects the modification of
the radial confinement, described by $R_{1}$ and $R_{3}$ in
Eq.~{\ref{eq:h1d}} due to magnetic field.   This approximation is valid for weak
magnetic fields $r\sub{c}\gg a$ that give the cyclotron
radius $r\sub{c}$ much larger than the ring radius $a$, as well as for the magnetic
fields of arbitrary strength confined to the interior of the ring. 

The requirement that the wave function of an orbiting hole is
single-valued, $\langle \phi = 2\pi|\Psi_{\kappa \tau}\rangle =
\langle \phi=0|\Psi_{\kappa \tau}\rangle$ gives the quantization
condition $\kappa = (2n+1)/2$, for integer $n$.  In the absence of
Zeeman coupling, the complete spectrum of the ring is periodic in the flux with the period
$\Phi_0$.  This perfect periodicity of the spectrum is broken by the
Zeeman interaction.

For the magnetic field in
$z-$ direction, normal to the plane of the ring, it is possible to
account exactly for the effects of Zeeman term $H\sub{Z} = bS_z$, where $b=
g_{zz} \mu\sub{B} B$ is the magnetic field in with absorbed Bohr magneton
$\mu\sub{B}$ and the gyromagnetic tensor component $g_{zz}$.  For 2DHG
the $g$-tensor is highly anisotropic, and to a good approximation the
only nonzero component is $g_{zz}$.  Therefore, this approximation is
valid also for the magnetic fields with in-plane components, with the
adjustment that $\vctr{B}\rightarrow (\vctr{B} \cdot
\vctr{e}_z)\vctr{e}_z$, since only the $z-$ component impacts both the
Aharonov-Bohm flux and the Zeeman term.

The Zeeman interaction couples the states of the same orbital momentum
$\kappa$ and opposite textures.  The energies and eigenstates
in the presence of Zeeman interaction are $( | \kappa \Uparrow \rangle
, | \kappa \Downarrow \rangle ) \rightarrow ( | \kappa + \rangle , | \kappa -
\rangle ) $ and $( E_{\kappa, \Uparrow} , E_{\kappa, \Downarrow} )
\rightarrow ( E_{\kappa, + } , E_{\kappa, -} )$, where 
\begin{equation}
\label{eq:ekappaplusminus}
\begin{split}
E_{\kappa, \pm} &= \frac{1}{2} \left( E_{\kappa, \Uparrow} + E_{\kappa,
  \Downarrow } \right) \pm 
\\
 & \sqrt{ \frac{1}{4} \delta (\kappa) ^2 + b^2 + b
  \cos\theta(\kappa) \delta (\kappa) }.
\end{split}
\end{equation}
The eigenstates in the presence of Zeeman interaction keep the
$\kappa$ quantum numbers, but the states of opposite textures get
mixed
\begin{equation}
\label{eq:statekappaplusminus}
\left(
\begin{array}{c}
| \kappa + \rangle 
\\
| \kappa - \rangle
\end{array}
\right)
=
\left(
\begin{array}{cc}
\cos \frac{\Theta(\kappa)}{2} &
- \sin \frac{\Theta(\kappa)}{2} 
\\
- \sin \frac{\Theta(\kappa)}{2} &
 \cos \frac{\Theta(\kappa)}{2}
\end{array}
\right)
\left(
\begin{array}{c}
| \kappa \Uparrow \rangle 
\\
| \kappa \Downarrow \rangle
\end{array}
\right),
\end{equation}

%
%
where the mixing angle $\Theta(\kappa)$ is 
\begin{equation}
\label{eq:mixingtheta}
\Theta(\kappa) = \arccos \frac{\frac{1}{2} \delta(\kappa) + b \cos \theta
  (\kappa)}{\sqrt{ \frac{1}{4} \delta(\kappa) ^2 + b^2 + b
  \cos\theta (\kappa) \delta(\kappa) }}.
\end{equation}
Here $\delta (\kappa)=E_{\kappa,\Uparrow}-E_{\kappa,\Downarrow}$ is
the energy difference of the two states with momentum $\kappa$ and
opposite textures.

\section{Tunneling model of conduction}

\begin{figure}[t]
\begin{center}
\includegraphics[width=8cm]{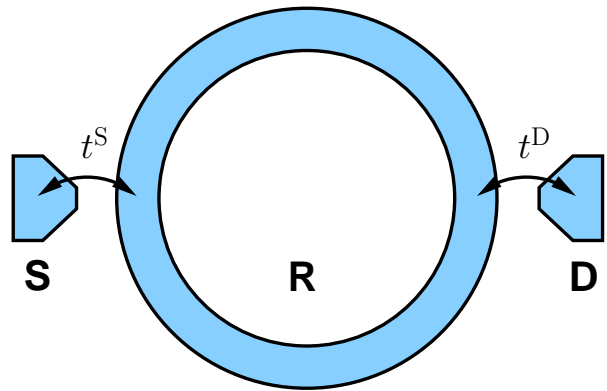}
\caption{\label{fig:ring} (Color online) Geometry of the ring of heavy
  holes coupled to a pair of leads.  The heavy holes in the ring (R)
  experience the spin-orbit coupling.  Due to this coupling, the
  eigenstates of holes confined to the ring have a spin texture.  The
  ring is coupled to the source (S) and drain (D) electrodes via
  tunneling of holes.  The tunneling is assumed to conserve the hole
  spin.}
\end{center}
\end{figure}

We consider a system of heavy holes confined to
a ring-shaped geometry and contacted by a pair of leads
[Fig. \ref{fig:ring}]. 
The lead density of states is assumed to be wide and flat.  In order
to elucidate the pseudospin structure of the leads, we allow for an
arbitrary pseudospin density matrix $\rho\subup{S(D)}$ in the
source(drain) lead.

The spin textures revealed in the eigenstates of heavy holes confined
to a ring influence the transport properties when the ring is coupled
to electrodes.  For example, the hole of a given pseudospin entering
the ring from the source electrode can propagate via different
$\Psi_{\kappa \tau}$ eigenstates, and arrive at the drain electrode
with different pseudospin orientations.  The pseudospin states at the
drain electrode will interfere, and the probability of transmission
will depend on the pseudospin orientations.  Since the pseudospin
orientations at the drain electrode depend on the hole momentum
$\kappa$ through the texture of the state $\Psi_{\kappa\tau}$, we may
expect that the transmission of the ring, and therefore the resulting
conductance will also depend on the momentum of the incoming hole.  This
momentum dependence is absent the electronic systems where the
texture is determined solely by the spin-orbit coupling constant
\cite{MPV04}.  Therefore, we expect new effects of spin interference
in transport of carriers that are subject to the cubic spin-orbit coupling.

The interference of heavy holes will be observable if their coherence
length $\lambda\sub{coh}$ is longer than the ring circumference
$\lambda\sub{ring}$.  At the same time, the spin-orbit length
$\lambda\sub{SO}$ that a hole must transverse in the ring in order
to experience an appreciable pseudospin rotation must be at least comparable
to $\lambda\sub{ring}$.  The resulting set of constraints
$\lambda\sub{coh} \gg \lambda\sub{ring}\sim \lambda\sub{SO}$ can be
achieved in the heavy hole structures based on carbon-doped GaAs
\cite{GLI+07}, \cite{YPS02}.

In order to find the transmission through the ring, we introduce a tunneling
Hamiltonian model for the ring coupled to source and drain electrodes.
The tunneling Hamiltonian description is valid when the overlap of the
electrode states and the ring states is small, $|\phi_{k\sigma}({\bf
  x})\Psi_{\kappa\tau}({\bf x})|\ll 1$ for every point ${\bf x}$
within the system, and every pair of states $(\kappa \tau , k \sigma)$.

The tunneling between either electrode and the ring occurs
on the length scale $\lambda\sub{tun}$ that is much shorter than
the spin-orbit length, $\lambda\sub{tun} \ll \lambda\sub{SO}$.
Therefore, unless there are magnetic impurities in the boundary 
region between the ring and the leads, the tunneling will preserve the
true hole spin, resulting in the hole pseudospin conservation in
tunneling, and the pseudospin independence of the tunneling amplitudes.

The tunneling Hamiltonian reads 
\begin{equation}
\label{eq:htun}
H\sub{T}= H\sub{S} + H\sub{D} + H\sub{R} + H\sub{T}, 
\end{equation}
where the three noninteracting Hamiltonians
\begin{align}
\label{eq:hs}
H\sub{S} &= \sum_{k\sigma} \epsilon_{k\sigma}\subup{S}
\cre{s_{k\sigma}} \ann{s_{k\sigma}}, 
\\
\label{eq:hd}
H\sub{D} &=
\sum_{k\sigma} \epsilon_{k\sigma}\subup{D}
\cre{d_{k\sigma}} \ann{d_{k\sigma}}, 
\\
\label{eq:hr}
H\sub{R} &=
\sum_{\kappa\tau} \epsilon_{\kappa\tau} \cre{r_{\kappa\tau}}
\ann{r_{\kappa\tau}}
\end{align}
describe decoupled source electrode, drain electrode, and the ring.
The operators $\ann{s_{k\sigma}}(\ann{d_{k\sigma}})$ annihilate a hole
of momentum $k$ and pseudospin $\sigma=\uparrow, \downarrow$ in the
source(drain) electrode, while the operators $\ann{r_{\kappa\tau}}$,
$\tau=\Uparrow,\Downarrow$ annihilates a hole in the ring state
$\Psi_{\kappa\tau}$.  The ring energies $\epsilon_{\kappa\tau}$ are
given by (\ref{eq:energiesDresselhaus}) and
(\ref{eq:energiesRashba}).  The tunneling term $H\sub{T}$ describes processes
when a hole hops from an electrode to the ring and back,
\begin{equation}
\label{eq:tunnelterm}
H\sub{tun}=\sum_{k\sigma,\kappa\tau} \left( t_{k\sigma,\kappa\tau}\subup{S}
\cre{s_{k\sigma}} \ann{r_{\kappa\tau}} +
t_{k\sigma,\kappa\tau}\subup{D} \cre{d_{k\sigma}} \ann{r_{\kappa\tau}}
+ \hc \right).
\end{equation}
The tunneling matrix elements, $t_{k\sigma,\kappa\tau}\subup{S(D)}$,
are determined by the details of the potential barrier between the
electrodes and the ring.  We are interested in the consequences of
nontrivial spin textures in the transport of holes through a ring.  The
potential barrier is due to electric fields, and its influence on the spin
and the hole pseudospin can come only from the spin-orbit coupling.
Here we assume that the holes of arbitrary pseudospin see the same
potential.  This assumption is valid for a potential which is nonzero
only in a tunneling region of the linear dimension much smaller than
the spin-orbit length.

Under these assumptions, we can model the tunneling matrix elements as
\begin{equation}
\label{eq:ts}
t_{k\sigma,\kappa\tau}\subup{S(D)} = t_{k,\kappa}\subup{S(D)} \langle
k\sigma {\rm S(D)} | \phi \sub{S(D)} \rangle \langle \phi\sub{S(D)} |
\kappa\tau\rangle,
\end{equation}
where the spin- and texture-independent matrix elements
$t_{k,\kappa}\subup{S(D)}$ describe the tunneling in the absence
of spin-orbit coupling, and the spin- and texture-dependent factor is
proportional to the overlap of the spin and texture part of the
wave function at the position $\phi\sub{S(D)}$ of the source (drain) junction.

The resulting tunneling Hamiltonian $H\sub{tun}$ is a generalization of
the Fano-Anderson model \cite{F61} to the many isolated
levels in a continuum with different couplings to the continuum states
in the leads.  Since the tunneling term $H\sub{T}$ in (\ref{eq:tunnelterm}) is
bilinear in the operators that describe the uncoupled system, it is in
principle exactly solvable.  However, the exact solution for the
eigenstates is simple and transparent only in the case of a single
level \cite{A93,EL04}.  The exact solution requires inversion of an
$N\times N$ matrix, where $N$ is the number of relevant ring states.
Instead of solving for the eigenstates, we calculate the current
through the ring using the Keldysh technique \cite{HJ96}.

The current through a region coupled to the leads via a tunneling
Hamiltonian was considered by Meir and Wingreen in \cite{MW92}.  Quite
generally, the current is 
\begin{equation}
\label{eq:i}
I=\frac{e}{h}\int \di \epsilon \left[ f\sub{S}(\epsilon) -
  f\sub{D}(\epsilon) \right] \tr \left[{\bf G}\subup{A}{\bf
  \Gamma}\subup{D}{\bf G}\subup{R}{\bf \Gamma}\subup{S}(\epsilon)\right],
\end{equation}
where $f\sub{S(D)}$ are Fermi distribution functions in the source and drain
electrodes, ${\bf G}\subup{R(A)}$ are retarded (advanced) Green
functions of the ring coupled to the leads, and ${\bf
  \Gamma}\subup{S(D)}$ are the escape rates of the ring states to
the source (drain) electrode.  The trace is taken over the ring
states $\kappa\tau$.  At zero temperature $T=0$, the differential conductance $g(\epsilon)$ for
the carriers of energy $\epsilon$ can be directly read off from
(\ref{eq:i}) (for finite temperature $T$, see below) as 
$
g(\epsilon) = \tr \left[ {\bf G}\subup{A} {\bf \Gamma}\subup{D} {\bf
    G}\subup{R} {\bf \Gamma}\subup{S} (\epsilon) \right]
$.

The Green functions in frequency space ${\bf G}\subup{R(A)}(\omega)$ are expressed in terms of
the self-energy as
\begin{equation}
\label{eq:green}
G\subup{R(A)}(\omega) =
\frac{1}{\left[{\bf g}\subup{R(A)}(\omega)\right]^{-1} - {\bf \Sigma}\subup{R(A)}(\omega)}.
\end{equation}
Here, ${\bf g}\subup{R(A)}$ is the retarded(advanced) Green function
of the ring.  In our noninteracting case, the self-energy ${\bf
  \Sigma}\subup{R(A)}$ is given exactly as a sum of contributions
coming from the excursion of the hole through the electrodes,
\begin{equation}
\label{eq:sigma}
\Sigma_{\kappa_1\tau_1,\kappa_2\tau_2}\subup{R(A)}(\omega) 
=
\sum_{k\sigma,{\rm L}}
\left(t_{k\sigma,\kappa_1\tau_1}\subup{L}\right)^*
g_{k\sigma}\subup{L~R(A)} (\omega) 
t_{k\sigma,\kappa_2\tau_2}\subup{L},
\end{equation}
where $g_{k\sigma}\subup{L~R(A)} (\omega)$ are retarded (advanced)
Green functions of decoupled leads, being diagonal in $k\sigma$.

The escape rates ${\bf{\Gamma}}\subup{S/D}$ describe the processes in
which a hole escapes from the ring into a lead and gets replaced by
another hole. They are defined as
\begin{equation}
\label{eq:escaperatesdef}
\Gamma_{\kappa_1 \tau_1,\kappa_2\tau_2}\subup{S/D}(\omega)  = 2\pi
\sum_{k\sigma} t\subup{S/D}_{k\sigma,\kappa_1\tau_1}
\left(t\subup{S/D}_{k\sigma,\kappa_2\tau_2}\right)^* \delta(\omega -
  \epsilon\subup{S/D}_{k\sigma}).
\end{equation}

The current through the ring is determined by Eqs.
(\ref{eq:i}), (\ref{eq:green}), (\ref{eq:sigma}), and (\ref{eq:escaperatesdef}), once we
incorporate the tunneling matrix elements (\ref{eq:ts}).  The current
will depend on the pseudospin states in the leads.  The effects of the
texture in the ring eigenstates will be visible in the conductance if
the states in the ring are polarized.  We thus consider general pseudospin
density matrices in the source (drain) electrode 
\begin{equation}
\label{eq:rholead}
\rho\subup{S(D)} = \frac{1}{2} \left( {\bf{1}} +
\vctr{P}\subup{S(D)}\cdot{\vectorsigma} \right),
\end{equation}
where the direction of $\vctr{P}\subup{S(D)}$, defines the axis of
partial polarization $|\vctr{P}\subup{S(D)}| \le 1$ in the
source(drain) lead.

We proceed by calculating the current using (\ref{eq:i}), with the
spin-dependent density of states in the escape rates
(\ref{eq:escaperatesdef}), and assuming that the bands in the leads
are wide and flat.  Our calculation is numerical and includes a finite
number $(184)$ of states in the ring.  This approach produces results
that do not change in the range of low values of $\omega$ with the
addition of new levels.  Another reason for truncating the number of
levels is the fact that the dispersion relations for heavy holes in
the ring (\ref{eq:energiesDresselhaus}), and (\ref{eq:energiesRashba})
predict unphysical states that are bound to the ring by strong
spin-orbit coupling.

\section{Differential conductance of a heavy-hole ring}

In this section, we discuss the influence of nontrivial pseudospin
textures in the eigenstates of the heavy hole ring to its
conductance.  In the tunneling picture, we can distinguish two basic
sources of the varying conductance.  One source is the discrete
spectrum of the ring, that in the limit of weak tunneling produces a
series of peaks in the conductance when the chemical potential of the
leads aligns with the discrete energy levels of the ring.  As we
increase the tunneling matrix elements the levels broaden due
to the coupling to the leads, and eventually begin to overlap.
Interference of the transitions from the source lead to the drain
lead via ring eigenstates is the second source of variations in the
conductance.  
\begin{figure}[t]
\includegraphics[width=8.5cm]{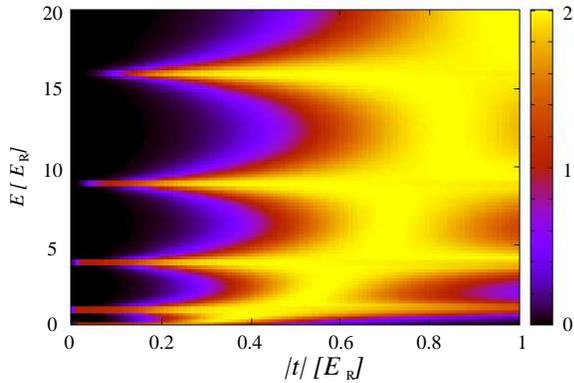}
\caption{(Color online) Tunneling dependence of differential conductance between
  unpolarized leads with Dresselhaus spin-orbit coupling in the
  leads.  The differential conductance $g(\epsilon)$ in units of the
  conductance quantum $G_0=h/e$ is plotted as a function of the absolute
  value of the tunneling matrix element between the states of
  uncoupled leads of the ring, and the chemical potential of the
  leads.  At small tunneling, the conductance shows peaks when the
  chemical potential of the ring aligns with the energy levels of the
  ring.  As the tunneling grows, the peaks become wider and begin to overlap.
   \label{fig:tabsPol0Dress}}
\end{figure}
\begin{figure}[t]
\begin{center}
\includegraphics[width=7cm]{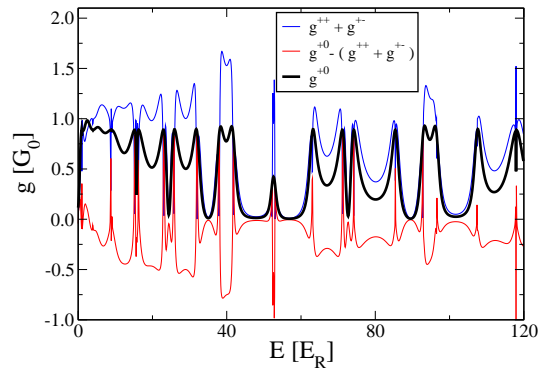}
\caption{(Color online) When the broadening of the ring levels is
  strong enough to produce the overlap of the energy levels, the
  tunneling processes through various states in the ring interfere.
  Pseudospin textures affect this tunneling.  The conductance $g^{+0}$
  between the pseudospin-polarized source lead and the unpolarized
  drain lead (thick black line) is not equal to the sum of
  conductances $g^{++}+g^{+-}$ between the polarized source and drain leads with
  parallel polarizations $g^{++}$ and the conductance between
  polarized source and drain lead with the antiparallel polarization
  $g^{+-}$ (thin dark (blue) line).  The difference $g^{+0}-(g^{++} +
  g^{+-})$ is the contribution of the interference term (thin bright
  (red)) line. 
\label{fig:spininterference}}
\end{center}
\end{figure}

We illustrate the interplay of these two mechanisms that modify
conductance by studying pseudospin-resolved current in the ring.
Then, we study the polarization-resolved conductance and show the qualitative
differences between Dresselhaus- and Rashba-coupled holes, which allow
for the determination of the dominant type of coupling.

Magnetic flux threaded through the ring causes Aharonov-Bohm
oscillations in the conductance that are further modified by the pseudospin textures.
The standard technique for observing these oscillations is by
looking for the peaks in the Fourier transform of the conductance as a function of
magnetic field that correspond to the period of one flux quantum.  We
show that the structure of Aharonov-Bohm oscillations in direct space,
i.e. before the Fourier transform, offers a signature of the cubic
spin-orbit coupling in the form of easily recognizable four-peak
structure in the oscillations.  We trace the emergence of this
split-peak structure to dependence of the energy spectrum of an orbiting
hole on the flux through the ring, and show that the form of the periodic
conductance is drastically different between the cubic and linear
spin-orbit coupling. 

The
possibility of experimental observation of the pseudospin-resolved
conductance is determined by the widths of the ring energy levels
compared to their splitting.  In our system, the levels broaden
due to tunneling.  In experiment, an additional thermal broadening
will further smear the conductance peaks.  We study the disappearance
of pseudospin-split conductance with temperature, and suggest the
regime favorable for resolving the pseudospin components.

In this section, the energy is measured in
units of $\unit{E_R}$, the energy of $\kappa=1$ orbital state in a
ring without spin-orbit coupling,
$\unit{E_R}=\hbar^2/2m\sub{hh}R_3^{-2/3}$.  For a typical ring of
radius $R_3^{-2/3}\sim 0.5\unit{\mu m}$, $\unit{E_R}\approx 1\unit{\mu eV}$.  

\subsection{Level broadening and interference}

The dependence of conductivity on the tunnel coupling strength and
carrier energy is illustrated in Fig.~\ref{fig:tabsPol0Dress} which
shows the conductance between unpolarized leads.  In the limit of
zero tunneling, $|t|\rightarrow 0$, the peaks in the conductance
appear at the energies of an isolated ring.  As the tunneling is
increased, the levels become broader, due to the tunneling of holes
between the ring and the lead.  Our calculation includes
contributions of an arbitrary number of such 'excursions'.  The
calculation is done at zero temperature (for the thermal broadening
see below).  With strong enough tunneling, the broadening of
the ring levels leads to their overlap.  The resulting conductance in
the overlapping region is not a simple sum of the conductances of
pseudospin components.  Since the tunneling involves many ring levels
in a coherent way, the resulting conductance shows a signature of
interference.  In Fig.~\ref{fig:spininterference}, we show
the interference term at a fixed tunneling strength. The conductance
$g^{+0}$ between the pseudospin-polarized source lead and the
unpolarized drain lead (thick black line) is not equal to the sum of
conductances $g^{++}+g^{+-}$ between the polarized source and drain
leads with parallel polarizations $g^{++}$ and the conductance between
polarized source and drain lead with the antiparallel polarization
$g^{+-}$ (thin dark (blue) line).  The difference $g^{+0}-(g^{++} +
g^{+-})$ is the contribution of the interference term (thin bright
(red)) line. 

\subsection{In-plane spin textures}

The conductance between the leads polarized in the direction normal to
the plane of the ring does not show the full difference between the
Dresselhaus- and Rashba coupling induced textures.  Namely the most
striking difference between the two textures is in the projection of
the pseudospin to the plane of the ring, see
Fig.~\ref{fig:textureDresselhaus} and Fig.~\ref{fig:textureRashba},
which is qualitatively different for the two forms of the
cubic spin-orbit coupling.  The in-plane component of the
Dresselhaus-only eigenstate winds once around the $z-$ axis as the
ring is transversed, and always stays tangential to the ring.  The
in-plane component of the Rashba state, on the other hand, winds three
times as the ring is transversed.

The winding of in-plane polarization is the same for all the states in the ring
and leaves a signature in the conductance.  We
calculate the conductance between the fully polarized leads with the
polarization vector $\vctr{P}$ in the plane of the ring, and with the
varying position of the drain lead along the ring.  We
notice that the conductance patterns in the Rashba case show more
islands of conductivity at a fixed carrier energy as the position of
the drain lead is encircling the ring.  The reason for the
additional islands is that the lead pseudospin aligns with the
in-plane projection of the pseudospin of ring eigenstates at the
position of the junction.  Aligned pseudospins increase the
conductivity and create the islands.  The in-plane projection of the
Dresselhaus eigenstate pseudospin texture aligns with lead polarization for one
junction position, while this alignment occurs for three positions in
the case of Rashba coupling. 
\begin{figure}[t]
\includegraphics[width=8cm]{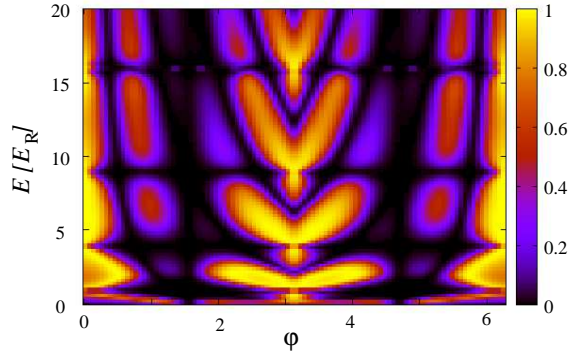}
\caption{Conductance between the completely in-plane polarized source
  and drain leads for Dresselhaus (upper panel) spin-orbit coupled
  holes. The position of the source lead is
  $\phi\sub{S}=0$, while the position of the drain lead
  $\phi\sub{D}$ varies between $0$ and $2\pi$.  For each drain
  position, the differential conductance is plotted as a function of
  the ring Fermi energy.  The radial structure of the pseudospin
  textures is seen in the traces of conductance at a fixed energy.
\label{fig:junpol1dressinplane}}
\end{figure}
\begin{figure}[t]
\includegraphics[width=8cm]{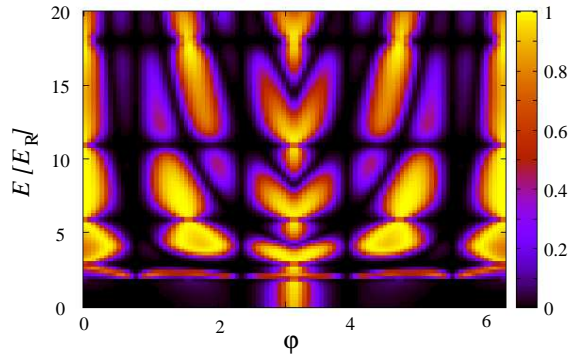}
\caption{Conductance between the completely in-plane polarized source
  and drain leads for Rashba spin-orbit coupled holes.  The position
  of the source lead is $\phi\sub{S}=0$, while the position of the drain lead
  $\phi\sub{D}$ varies between $0$ and $2\pi$.  For each drain
  position, the differential conductance is plotted as a function of
  the ring Fermi energy.  Compare with the case of Dresselhaus
  spin-orbit coupling.
\label{fig:junpol1rashinplane}}
\end{figure}

\subsection{Modified Aharonov-Bohm oscillations}

Conductance measurements between the polarized leads and with the
control over the chemical potential of the ring are difficult to
achieve.  Typical experiments measure the conductance as a function of the
magnetic field that
threads a magnetic flux through the ring and introduces the Zeeman
coupling.  In our model of tunneling conductance the Aharonov-Bohm
phase can be incorporated in the boundary conditions for the ring wave
function, using the singular gauge.  This leads to a quantization
condition for $\kappa - \Phi/\Phi_0$, where $\Phi$ is the flux
threaded through the ring, and $\Phi_0$ is the flux quantum.  The
effect of the flux is thus the shift of all the $\kappa$ quantum
numbers.  As a consequence, the energy levels and the pseudospin
textures change.  The new texture angles $\theta\subup{d/r}$ and the
new energies $E_{\kappa,\tau}$ are still given by Eqs.
(\ref{eq:energiesDresselhaus}), (\ref{eq:textureAngleDresselhaus}),
(\ref{eq:energiesRashba}), and (\ref{eq:textureAngleRashba}), but with the
shifted values of the orbital quantum number $\kappa \rightarrow
\kappa + \Phi/\Phi_0$.  

\begin{figure}[t]
\begin{center}
\includegraphics[width=7cm]{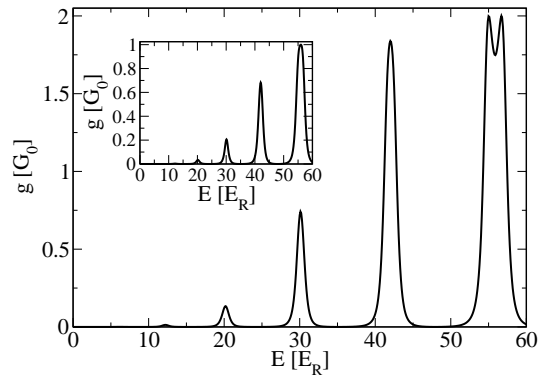}
\caption{Conductance of the ring threaded by a half of flux quantum.
  Leads are unpolarized.
  Note that the conductance is not zero due to tunneling through
  off-resonant states.  The inset shows the
  conductance of the same ring with the same flux, but between
  polarized leads.  The peak at $\epsilon\sub{F}\approx 55
  \unit{E_R}$ shows that the split peak in the main plot is due to
  the conductance of the holes of different polarizations. 
\label{fig:half_flux_pol}}
\end{center}
\end{figure}

The gross features of the Aharonov-Bohm oscillations can be
understood in terms of a simplified picture based on interference of
levels that lie close in energy.  The spectra of the ring in zero
magnetic field, and in the presence of weak spin-orbit coupling consists of pairs of closely spaced Kramers doublets
$(\Psi_{\kappa,\Uparrow}, \Psi_{-\kappa,\Downarrow})$ and
$(\Psi_{\kappa+1,\Downarrow},\Psi_{-\kappa-1,\Uparrow})$.  The gap
between these doublets scales as $\beta^2$ ($\alpha^2$) for weak
Dresselhaus (Rashba) spin-orbit coupling, while all the other states
are separated by larger gaps that originate from the kinetic
energy terms and persist in the absence of spin-orbit coupling.
Therefore, we can approximately describe the conductance by transition
amplitudes 
\begin{equation}
\label{eq:tamplitudes}
T=
\left(
\begin{array}{cc}
T^{+,+}  &  T^{+,-}
\\
T^{-,+} & T^{-,-}
\end{array}
\right),
\end{equation}
where the matrix element
$T^{s_1 s_2}$ stands for the amplitude for a hole of pseudospin $\pm
1/2$ for $s_1=\pm$
in the source-lead to tunnel into the drain-lead with the pseudospin
$\pm 1/2$ for $s_2=\pm$.  Taking into
account only the tunneling through the four closely spaced levels and
in the absence of the flux through the ring, the transition amplitudes are
\begin{equation}
\label{eq:4levels0flux}
T_0 = 
2\sin\left(\kappa\pi\right) 
\cos\left(\frac{d_0}{2}\right)
\left(
\begin{array}{cc}
\cos \frac{s_0}{2}
&
i\sin \frac{s_0}{2}
\\
i \sin \frac{s_0}{2}
&
\cos \frac{s_0}{2}
\end{array}
\right),
\end{equation}
where $s_0=\theta\subup{d/r}(\kappa)+\theta\subup{d/r}(\kappa + 1)$, and
$d_0=\theta\subup{d/r}(\kappa)-\theta\subup{d/r}(\kappa + 1)$ are the
sum and the difference of the texture angles of the involved states.
Similar considerations for the case of a ring threaded by the magnetic
flux $\Phi=\Phi_0/2$ equal to half the flux quantum gives
\begin{equation}
\label{eq:4levelshalfflux}
T_{1/2} = 
%
%
2\cos\left(\kappa\pi\right) 
\cos\left(\frac{d_{1/2}}{2}\right)
\left(
\begin{array}{cc}
\cos \frac{s_{1/2}}{2}
&
i\sin \frac{s_{1/2}}{2}
\\
i \sin \frac{s_{1/2}}{2}
&
\cos \frac{s_{1/2}}{2}
\end{array}
\right),
\end{equation}
where the relevant sums are now
$s_{1/2}=\theta\subup{d/r}(\kappa+1/2)+\theta\subup{d/r}(\kappa +
3/2)$, and
$d_{1/2}=\theta\subup{d/r}(\kappa+1/2)-\theta\subup{d/r}(\kappa +
3/2)$.  The quantum number $\kappa$ is a half of an odd integer and
$T_{1/2}=0$.  Therefore, this simplified description correctly predicts the
minima in conductance when half a flux quantum threads the ring.  The
conductance value is zero in this simple model, but it turns out to be
nonzero when the additional levels are included in the more detailed
model.  When the additional levels in the ring are included, the
conductance can be nonzero in the ring threaded by half of flux
quantum, see Fig.~\ref{fig:half_flux_pol}.
%
%
The currents transmitted through the ring carry hole polarization, as can be
seen from the figures.  The peak in the unpolarized conductance near
the energy $\epsilon=54\unit{E_R}$ is split, while the polarized
conductance shows a single peak of roughly half the height.  The
components of the split peak correspond to pseudospin components with
high polarization up and down, described by pseudospin density
matrices (\ref{eq:rholead}) with $\vctr{P}\approx \vctr{e}_z$ and
$\vctr{P}\approx -\vctr{e}_z$.  This splitting is a clear signature of
pseudospin dependent transport.  

The standard setup for a study of conductance oscillations as a
function of the magnetic field consists of measuring the
conductance at a fixed lead chemical potential and sweeping the
external magnetic field.  The conductance then typically reveals the
oscillations with the period $T\sub{AB}=S\Phi_0^{-1}$, $S$ being the
ring surface area and $\Phi_0$ the flux quantum.  The
spin-orbit coupling was found to modify these oscillations
\cite{GLI+07}.  In our model the conductance is modified due to the
presence of four closely spaced energy levels that correspond to each
peak in the conductance.  At zero flux these four levels are the Kramers
doublets $(\Psi_{\kappa,\Uparrow},\Psi_{-\kappa,\Downarrow})$ and
$(\Psi_{\kappa+1,\Downarrow},\Psi_{-(\kappa+1),\Uparrow})$.  The
splitting between these pairs in the absence of magnetic field is of second order in spin-orbit
coupling.  As the magnetic flux is threaded through the ring the
quartet of levels splits, with two of the levels with $\kappa>0$
gaining energy, and the levels with $\kappa<0$ losing it.  In addition
the Zeeman coupling splits these levels further.  This behavior is in
sharp contrast to the linear spin-orbit coupling case where there are
at most two states of any given energy.  

\begin{figure}[t]
\includegraphics[width=8.5cm]{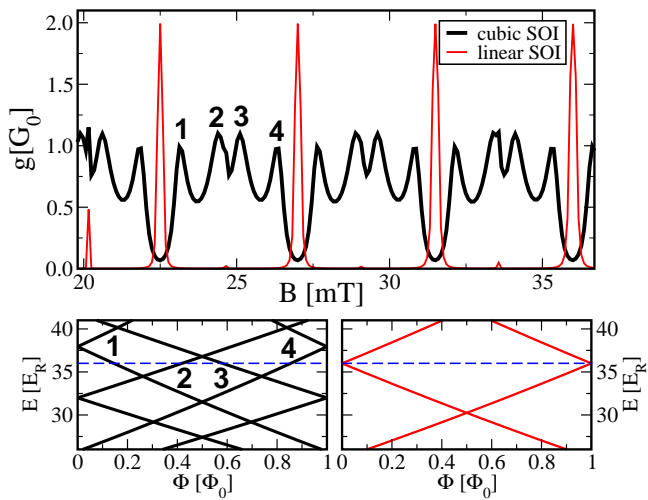}
\caption{Aharonov-Bohm oscillations for different types of spin-orbit
  coupling.  (a) The conductance of the ring as a function of the
  magnetic field shows oscillations with the period that corresponds
  to a flux quantum threading the ring for both linear ((red) light)
  and cubic ((black) dark) spin-orbit coupling, but with markedly
  different conductance within a period. (b,c) The four-peak structure
  (labels $1$-$4$ in (a)) for the cubic spin-orbit interaction, and
  the single-peak structure for the linear spin-orbit coupling can be traced to the
  magnetic fields at which an energy level in the ring aligns with the
  leads (labels $1$-$4$ in (b)).  Calculations for both plots are done
  for the lead chemical potential of $36 \unit{E_R} \approx 36
  \unit{\mu eV}$, close to an energy level of an isolated ring in the
  absence of spin-orbit coupling, and the linear spin-orbit coupling model is derived
  from the cubic one by setting $R_3=0$. 
\label{fig:directcubiclinear}}
\end{figure}

The four-peak structure within the maximum of conductance in
Aharonov-Bohm oscillations represents a
signature of the cubic spin-orbit coupling, see
Fig.~\ref{fig:directcubiclinear}.  The period of oscillations is equal for
both types of coupling, but the shape of the peaks is drastically
different.  
The four-peak structure is most visible when the leads are tuned into
the vicinity of a ring energy level.  At these energies, in contrast,
the linear spin-orbit coupling produces a single-peak structure.  

Fourier spectra of conductance fluctuations were
reported to show the signature of spin-orbit coupling in the diffusive
regime, seen in the splitting of peaks in the Fourier spectrum
\cite{GLI+07,HTS07}.  We have compared the Fourier spectra of our results in
the case of linear and cubic form of spin-orbit coupling.  In our
tunneling model, the Fourier spectra of the ring with linear
spin-orbit coupling differs from the spectra of the ring with the
cubic spin-orbit coupling in the relative size of
the base and higher harmonic.  The shape of the peaks in Fourier
spectrum does not show significant differences.
\begin{figure}[t]
\begin{center}
\includegraphics[width=6.78cm]{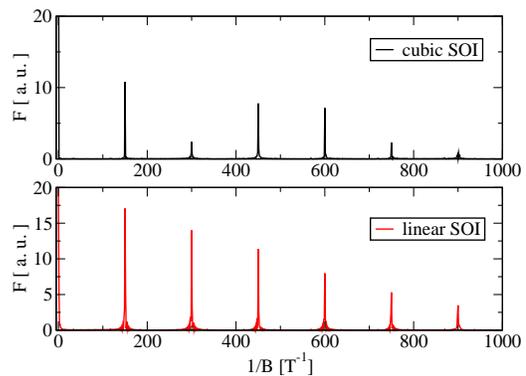}
\caption{Fourier spectra of the ring conductance as function of the
  magnetic field.  The upper panel ((black) dark) shows the conductance spectrum of
  the ring with cubic spin-orbit coupling, and the lower panel
  ((red) light) shows the conductance spectrum of the ring with linear spin-orbit
  coupling.  The ring radius is set to $R_3^{-3/2}=0.5 \unit{\mu m}$,
  and the lead chemical potential is $36 \unit{E_R}$.  The structure
  of base frequency and the higher harmonics is the
  consequence of the Aharonov-Bohm oscillations.  The two cases can be
  distinguished by the relative size of the harmonics.
\label{fig:fouriercubiclinear}}
\end{center}
\end{figure}
Therefore, the signature of the cubic spin-orbit coupling is clearly
visible in the direct-space Aharonov-Bohm oscillations, and very hard
to discern in the Fourier transform.

\subsection{Thermal broadening}

\begin{figure}[t]
\includegraphics[width=8cm]{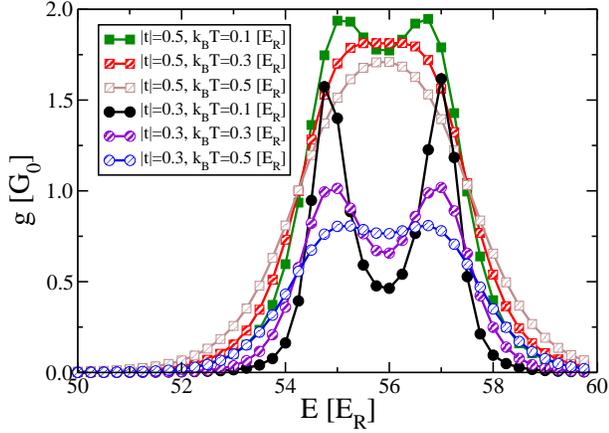}
\caption{Thermal broadening of differential conductance $g$.
  Components of the split peak in the
  differential conductance (inset in Fig.~\ref{fig:half_flux_pol})
  merge into a single
  peak as the temperature is raised.  The squares represent the
  conductance of the ring with stronger tunneling
  between the leads and the ring, $|t|=0.5 E\sub{R}$, while the
  circles represent the conductivity for weaker tunneling $|t|=0.3
  E\sub{R}$.  Weaker tunnel coupling allows the splitting to be
  resolved at higher temperatures. 
\label{fig:thermal}}
\end{figure}

The split peaks in differential conductance as function of lead
chemical potential will be visible is the distance between the peaks
is larger then their width.  As an illustration of the effects of
temperature $T>0$, we will investigate the broadening of
pseudospin-resolved peak at half flux quantum $\Phi= \Phi_0/2$,
Fig.~\ref{fig:thermal}.  For the parameters we used, the splitting of the peaks is $\sim
3\unit{\mu eV}\approx 30\unit{m K}$, and requires low temperatures
to resolve.  The broadening that impairs resolving of the split peaks
has a temperature-independent contribution due tunneling to the leads,
and it is further increased due to the temperature.  We study the thermal broadening of the conductivity
using (\ref{eq:i}), and finding the conductance $g$ at finite
temperatures.  We find that the conductance is indeed broadened at
finite temperatures, Fig.~\ref{fig:thermal}.  However, the visibility of the peaks and the resolution
of peaks can be improved if the peaks are narrower or the
splitting is larger.  The peak splitting grows with the absolute value
of the momentum, $|\kappa|$, and can be observed at higher temperatures if the momentum of
the interfering states is larger.  In summary, the favorable
conditions for the observation of pseudospin- dependent conductance
are weak tunneling and low temperatures.  Both of these conditions aim at
reducing the line width of the peaks.  Another way to resolve the
pseudospins is to perform an experiment with the higher chemical
potential in the leads, and observe the splitting of the higher-energy
peak.  These peaks are further separated in energy, due to the cubic
spin orbit coupling.   

\section{Conclusions}

We have investigated the conductance of a mesoscopic ring of heavy
holes tunnel-coupled to leads.  In the coherent regime, the
transport through the ring is dominated by the the energy spectrum and the
pseudospin texture of the orbiting hole eigenstates.  Due to
the cubic form of spin-orbit interaction,  the pseudospin texture of
the hole eigenstates is momentum dependent, as opposed, e.g., to the
electrons with linear spin-orbit coupling.  

The hole transport proceeds through tunneling between the source and
the drain lead via various ring eigenstates, with the phase of each tunneling path
modified due to the spin texture.  The effects of interference between
the tunneling paths are visible in the conductance when the tunnel
broadening is sufficient to make the ring energy levels overlap.  We
have demonstrated that the dominant type of spin-orbit interaction can
be deduced from the pseudospin-dependent conductance between the
polarized leads. 

Aharonov-Bohm oscillations appear in the tunneling approach as a
consequence of the evolution the ring spectrum as the magnetic flux is
threaded through the ring.  Approximately periodic evolution of the
peaks leads directly to the approximately periodic conductance
oscillations.  We have explained the four-peak shape of the Aharonov-Bohm
oscillations in the direct space as a direct consequence of four-fold near
degeneracy of the orbiting hole energy levels.  This particular shape
of Aharonov-Bohm oscillations is a signature of the cubic spin-orbit
coupling, but it is not visible in the Fourier transform of the conductance.

The pseudospin splitting of the conductance peaks, caused by pseudospin
textures of the ring eigenstates is clearly visible at zero
temperature and low tunneling, but disappears when the combined
thermal- and tunnel broadening becomes comparable to the size of the splitting.

We thank M. Trif, J. C. Egues, and C. Bruder for useful
discussions.  We acknowledge financial
support from the Swiss SNF, the NCCR Nanoscience Basel; the EU under
``MagMaNet'', and ``MolSpinQIP'', Brain Korea 21 Project, DFG within SPP 1285 ``Spintronics'' and FOR 921.


\begin{thebibliography}{32}

\bibitem{AB59}
Y.~Aharonov and D.~Bohm, Phys. Rev. \textbf{115}, 485 (1959).

\bibitem{AC84}
Y.~Aharonov and A.~Casher, Phys.\ Rev.\ Lett. \textbf{53}, 319 (1984).

\bibitem{LC04}
M.~Lee and M.~Y. Choi, Journal of Physics A: Mathematical and General
\textbf{37}, 973 (2004).

\bibitem{BLK+08}
M.~F. Borunda, X. Liu, A.~A. Kovalev, X.-J. Liu, T. Jungwirth, and
J.~Sinova, arXiv:0809.0880  (2008).

\bibitem{BKS+06}
T.~Bergsten, T.~Kobayashi, Y.~Sekine, and J.~Nitta,
Phys.\ Rev.\ Lett. \textbf{97}, 196803 (2006).

\bibitem{PZ08}
  M.~Pletyukhov and U.~Z\"{u}licke, Phys.\ Rev.\ B \textbf{77}, 193304 (2008).

\bibitem{UWL+84}
C.~P. Umbach, S.~Washburn, R.~B. Laibowitz, and R.~A. Webb,
Phys.\ Rev.\ B \textbf{30}, 4048 (1984).

\bibitem{EIA03}
O.~Entin-Wohlman, Y.~Imry, and A.~Aharony,
Phys.\ Rev.\ Lett.\textbf{91}, 046802 (2003).

\bibitem{LGB90}
 D.~Loss, P.~Goldbart, and A.~V. Balatsky,
 Phys.\ Rev.\ Lett. \textbf{65}, 1655 (1990).

\bibitem{LG92}
D.~Loss and P.~M. Goldbart, Phys.\ Rev.\ B \textbf{23}, 13544 (1992).

\bibitem{NMT99}
J.~Nitta, F.~E. Meijer, and H.~Takayanagi,
App.\ Phys.\ Lett.\textbf{75}, 695 (1999). 

\bibitem{CRM06}
R.~Citro, F.~Romeo, and M.~Marinaro, Phys.\ Rev.\ B \textbf{74}, 115329 (2006).

\bibitem{CR07}
R.~Citro and F.~Romeo, Phys.\ Rev.\ B \textbf{75}, 073306 (2007).

\bibitem{CR08}
R.~Citro and F.~Romeo, Phys.\ Rev.\ B \textbf{77}, 193309 (2008).

\bibitem{GLI+07}
B.~Grbic, R.~Leturcq, T.~Ihn, K.~Ensslin, D.~Reuter, and A.~D. Wieck,
Phys.\ Rev.\ Lett. \textbf{99}, 176803 (2007).

\bibitem{HTS07}
B.~Habib, E.~Tutuc, and M.~ Shayegan, App.\ Phys.\ Lett. \textbf{90},
152104 (2007).

\bibitem{PGK+04}
M.~G. Pala, M.~Governale, J.~K{\"o}nig, and U.~Z{\"u}licke,
Europhys. Lett. p. 850 (2004).

\bibitem{EL04}
H.-A. Engel and D.~Loss, Phys.\ Rev.\ Lett. \textbf{93}, 136602 (2004).

\bibitem{KBJ+07}
A.~A. Kovalev, M.~F. Borunda, T. Jungwirth, L.~W. Molenkamp, and
J.~Sinova, Phys.\ Rev.\ B \textbf{76}, 125307 (2007).

\bibitem{SN04}
S.~Souma and B.~K. Nikoli{\'c}, Phys.\ Rev.\ B \textbf{70}, 195346 (2004).

\bibitem{HKS01}
W.~{Hofstetter}, J.~K\"onig, and H.~Schoeller,
Phys.\ Rev.\ Lett. \textbf{87}, 156803 (2001).

\bibitem{KG02}
J.~K\"onig and Y.~Gefen, Phys.\ Rev.\ B \textbf{65}, 045316 (2002).

\bibitem{SEA05}
P.~Simon, O.~Entin-Wohlman, and A.~Aharony,
Phys.\ Rev.\ B \textbf{72} (2005).

\bibitem{LSG93}
D.~Loss, H.~Schoeller, and P.~M. Goldbart, Phys.\ Rev.\ B \textbf{48},
15218 (1993).

\bibitem{YPS02}
Y.-B. Yau, E.~P. De~Poortere, and M.~Shayegan,
Phys.\ Rev.\ Lett. \textbf{88}, 146801 (2002). 

\bibitem{BL05}
D.~V. Bulaev and D.~Loss, Phys.\ Rev.\ Lett. \textbf{95}, 076805 (2005).

\bibitem{MMK02}
F.~E. Meijer, A.~F. Morpurgo, and T.~M. Klapwijk, Phys.\ Rev.\ B
\textbf{66}, 033107 (2002).

\bibitem{MPV04}
B.~Moln\'{a}r, F.~M. Peeters and P.~Vasilopoulos, Phys.\ Rev.\ B
\textbf{69}, 155335 (2004).

\bibitem{F61}
U.~Fano, Phys. Rev. \textbf{124}, 1866 (1961).

\bibitem{A93}
D.~Averin, J. Appl. Phys. \textbf{73}, 2593 (1993).

\bibitem{HJ96}
H.~Haug and A.-P. Jauho,
  \emph{Quantum Kinetics in Transport and Optics of Semiconductors} (Springer, 1996).

\bibitem{MW92}
Y.~Meir and N.~S. Wingreen, Phys.\ Rev.\ Lett. \textbf{68}, 2512 (1992).

\end{thebibliography}
\end{document}